# Research Output of Webology Journal (2013-2017): A Scientometric Analysis

## Muneer Ahmad[1], Dr. M. Sadik Batcha[2], Basharat Ahmad Wani[3], Mohammad Idrees Khan[4] & S. Roselin Jahina[5]

[1,5]*Ph.D Research Scholar, Department of Library and Information Science,*
*Annamalai University, Tamil Nadu, India*
[2]*Associate Professor, Department of Library and Information Science,*
*Annamalai University, Tamil Nadu, India*
[3, 4] *Library Professional, Central University of Kashmir, Nowgam Srinagar, Jammu & Kashmir, India*

## ABSTRACT

*Webology is an international peer-reviewed journal in English devoted to the field of the World Wide Web and serves as a forum for discussion and experimentation. It serves as a forum for new research in information dissemination and communication processes in general, and in the context of the World Wide Web in particular. This paper presents a Scientometric analysis of the Webology Journal. The paper analyses the pattern of growth of the research output published in the journal, pattern of authorship, author productivity, and subjects covered to the papers over the period (2013-2017). It is found that 62 papers were published during the period of study (2013-2017). The maximum numbers of articles were collaborative in nature. The subject concentration of the journal noted was Social Networking/Web 2.0/Library 2.0 and Scientometrics or Bibliometrics. Iranian researchers contributed the maximum number of articles (37.10%). The study applied standard formula and statistical tools to bring out the factual result.*

***Keywords****: Author Productivity, Bibliometrics, Collaboration pattern, Iran, Scientometrics, Webology.*

## I. INTRODUCTION

Webology is an international peer-reviewed academic journal in English devoted to the field of the World Wide Web and serves as a forum for discussion and experimentation. The year of inception was 2004. Its frequency was quarterly from 2004-2008 and started biannual from 2009 onwards. It serves as a forum for new research in information dissemination and communication processes in general, and in the context of the World Wide Web in particular. Concerns include the production, gathering, recording, processing, storing, representing, sharing, transmitting, retrieving, distribution, and dissemination of information, as well as its social and cultural impacts. There is a strong emphasis on the Web and new information technologies. Special topic issues are also often seen, Noruzi, A. (2016)[1]. The journal is listed in the online catalogues and directories of open access journals of several prestigious university libraries around the world. This journal is indexed by the following services:

Scopus: Elsevier Bibliographic Databases,ProQuest ,EBSCO,LISA: Library & Information Science Abstracts,DOAJ - Directory of Open Access Journals, Open J-Gate,FRANCIS,Web Citation Index, Academic Journals Database, China Education Publications Import & Export Corporation (CEPIEC).

Scientometrics is the study of the quantitative aspects of the process of science as a communication system. It is centrally, but not only, concerned with the analysis of citations in the academic literature. In recent years it has come to play a major role in the measurement and evaluation of research performance. In this review we consider: the historical development of scientometrics, sources of citation data, citation metrics and the "laws" of scientometrics, normalization, journal impact factors and other journal metrics, visualizing and mapping science, evaluation and policy, and future developments. Scientometrics – "The quantitative methods of the research on the development of science as an informational process" Nalimov & Mulcjenko (1971)[2]. This field concentrates specifically on science (and the social sciences and humanities).The field of library and information science (LIS) has developed several quantitative methods to study the various aspects of subjects. The different metrics of LIS are continuously increasing, starting from librametrics, bibliometrics, scientometrics, informatics, webometrics, geometrics to cybernetics Khan (2016)[3] cited in Singh et al. (2017)[4]. The present study explores the blueprints of scholarly communication of Webology journal from the year 2013-2017 and determines to exposure the quality of contributions of this journal towards library and information science literature.

## II. REVIEW OF RELATED LITERATURE

Khan (2016)[3] studied and reported that majority of the authors preferred journals as an information source for writing of scholarly communication. It was suggested by the author that the journal should try to get high-quality papers from foreign authors too, which may be useful in enhancing its global impact and reputation. Singh et al. (2017)[5] carried analysis of 283 research articles of international journal of library and information studies (IJLIS) during the period 2012-2016. It was clear from the findings that only four different countries across the world have dominating this journal during the period of study. In similar study conducted by, Varma and Singh (2017)[6] claimed that from 2013 to 2016 articles publication rate have been increased and conference papers were the most widely used form of documents in which most of the literature on the subject 'big data' has been published and observed various aspects e.g. year wise distribution of article, authorship pattern, most prolific authors, geographical distribution of authors, bibliographic form used for citations and length of article etc. After analyzing bibliographic forms 3685 references were found in the 283 articles. It was clear that only four different countries across the world have contributed research articles to this journal during the period of study. M.Sadik Batcha and Muneer Ahmad (2017)[7] did comparative analysis of Indian Journal of Information Sources and Services (IJISS) and Pakistan Journal of Library and Information Science (PJLIS) during 2011-2017 and studied various aspects like year wise distribution of papers, authorship pattern & author productivity, degree of collaboration pattern of Co-Authorship , average length of papers , average keywords,etc and found 138(94.52%) of contributions from IJISS were made by Indian authors and similarly 94(77.05) of Contributions from PJLIS were done by Pakistani authors. Papers by Indian and Pakistani Authors with Foreign Collaboration are minimal (1.37% of articles) and (4.10% of articles) respectively. Suresh et al. (2015)[8] reported that 97.33 % of the papers were published by multi author. It was revealed that the Growth rate is 0.41 in 2010 and which decreased up to 0.19 in 2014 and most of the articles contributed from India. Thanuskodi (2010)[9] examined the research output of social scientists on social science subjects. The study cover year wise, institution-wise, country-wise, authorship pattern, range of references cited of the articles etc. On the other hand, Velmurugan (2013)[10] examined the research output of 203 article s appear in Annals of Library and Information Studies journal. It was found that the most of the contributions are co-authored 88 (43.35 %.). The degree of collaboration ranges from 057 to 0.82 and the average degree of collaboration is 0.64. The total average number of authors per paper is 1.87 and the average productivity per author is 0.53. On the other hand, Singh (2012)[11] studied and shows that maximum numbers of contributions are single author with 124 papers (56.10%). It was also clear that Indian contributions in this journal are significantly less (1.87%). Hussain and Fatima (2011)[12] explained that the majority of the articles were contributed by single authors. It was also clear that authors were librarians, faculty members and researchers associated with academic and research organization. Rajendran et al. (2011)[13] scrutinized the 633 research articles published in Journal of Scientific and Industrial Research (2005-2009) and revealed that the highest number of research papers contributed by multiple authors during the study period. It was also clear from the study that the degree of collaboration was 0.92. Sanni and Zainab (2010)[14] studied the scholarly communication published in Medical Journal of Malaysia during 2004-2008 and found 28(4.82%) of contributions were made by Malaysian authors with foreign collaboration. Batcha et al.(2018)[15] did scientometric analysis of the DESIDOC Journal and analyzed the pattern of growth of the research output published in the journal, pattern of authorship, author productivity, and, subjects covered to the papers over the period (2013-2017). It found that 227 papers were published during the period of study (2001-2012). The maximum numbers of articles were collaborative in nature. The subject concentration of the journal noted was Scientometrics. The maximum numbers of articles (65 %) have ranged their thought contents between 6 and 10 pages. M.Sadik Batcha and Muneer Ahmad (2017)[16] conducted scientometric analysis of 146 research articles published in Indian journal of Information Sources and Services (IJISS). The number of contributions, authorship pattern & author productivity, average citations, average length of articles, average keywords and collaborative papers was analyzed. Out of 146 contributions, only 39 were single authored and rest by multi authored with degree of collaboration 0.73 and week collaboration among the authors. The study concluded that the author productivity was 0.53 and was dominated by the Indian authors. Serenko et al. (2009)[17] described that an average manuscript was written by 1.73 authors. The USA, Canada and the UK were the three most productive countries, which is consistent with prior KM/IC productivity research.

**Objectives**

Five volumes of Webology (2013-2017) comprising of 10 issues have been studied to find out.

➢ Year wise distribution of contributions in the journal.

➢ Subject wise distribution of papers.

➢ Single and joint contributions.

➢ Global distribution of contributions.

- ➢ Author productivity.
- ➢ Number of citations used.
- ➢ Average number of citations per volume.
- ➢ Institution wise distribution of contributions.
- ➢ Degree of Collaboration in Webology journal.

## III. METHODOLOGY

The earlier study on Webology was conducted by Chandran Velmurugan and Radhakrishnan Natarajan (2015)[18]. They have studied Volume 04 to 10 (2007 - 2013). The present study is an effort to make it update by studying Volume 10 to 14 (2013-2017).The research papers given in these 5 volumes of Webology Vol. 10 - 14 (2013-2017) have been studied in this paper. The analysis includes 62 research articles. A data sheet was created on different aspects for main articles. The data were collected from the website of Webology Journal. The study uses appropriate measures and techniques of scientometric analysis. Keeping the objectives of the study in mind, the data were collected from the Webology Journal covering the 5 identified years. The analysis covers mainly the number of articles published per volume in each of the specified years, the authorship patterns, the subject areas covered, the length of articles, the citation pattern of articles and the article types.

## IV. ANALYSIS AND DISCUSSION

## 1. Chronicle Distribution

The collected data of 62 research articles covered in five volumes from 10 to 14 have been analyzed using the statistical tools. There have been two issues found in every volume and totally 10 issues have been taken for analysis. Cambridge Dictionary has elaborated the term Chronology which means, a list or explanation of events in the order in which they happened. This assist the researcher to scrutinize the gradual increase or decrease of the growth of literature output in a particular subject. It also helps the researcher to predict in which year more articles were published on a particular subject. Figure 1 depicts the chronicle growth rate of research outputs produced by different authors in Webology during 2013- 2017. It could be observed that out of 62 articles, the maximum number 15 (24.19%) of scholarly articles were published in 2014 whereas the minimum number 11 (17.74%) of articles were in 2017.

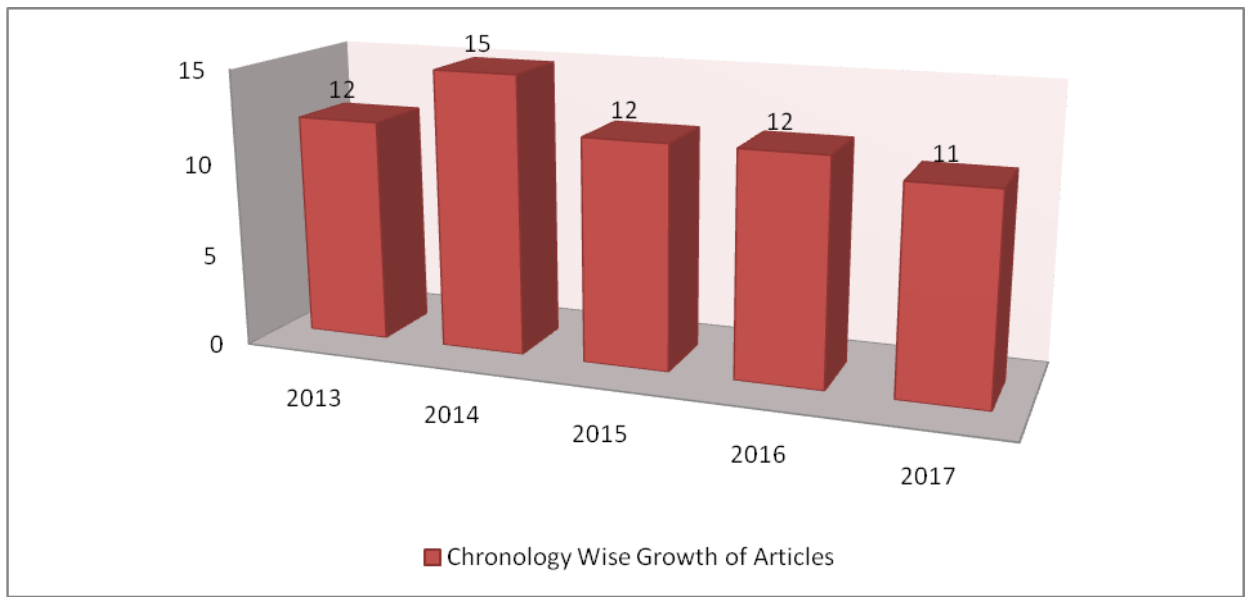

**Figure 1: Chronicle wise Growth of Articles**

### 2. Authorship Pattern of Single vs Joint Authors

Figure 2 shows the details about the single and joint-authored papers. A total of 18 contributions (29.03%) have been contributed by single authors, 44 contributions (70.97 %) by joint authors. It shows that the highest number of contributions were joint authored papers.

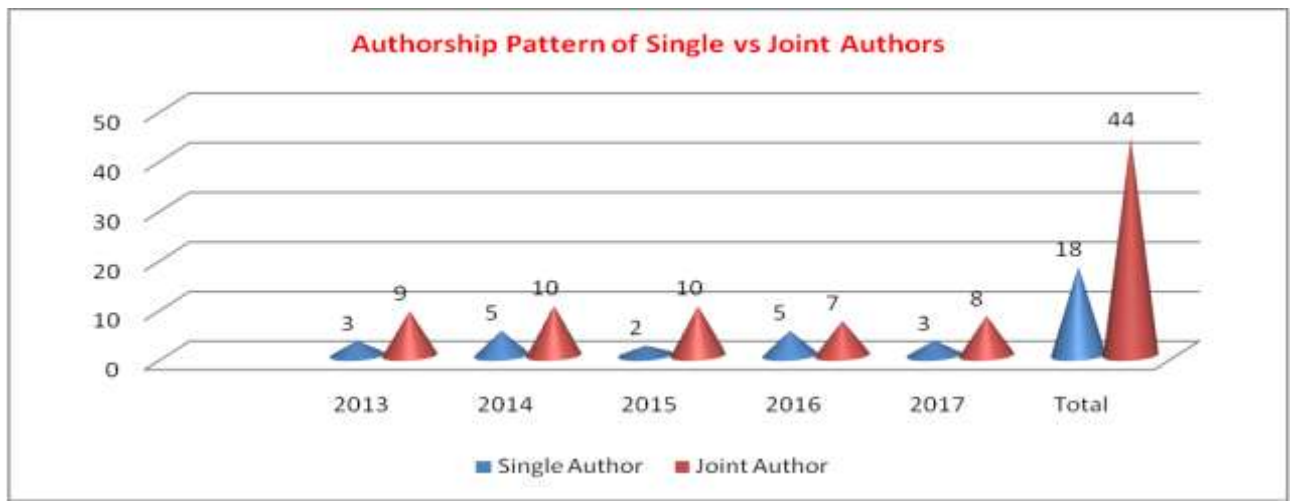

**Figure 2: Single vs Joint Authorship Pattern**

### 3. Year - wise Single vs Multi Authorship Pattern

Table 1 indicates the year wise authorship pattern linking with single versus multi authored research output. In this regard, the highest number 44(70.97%) of papers were contributed by multi authors whereas the remaining 18(29.03%) papers were produced by single authors. It shows the maximum number of contributors were multi authored papers. Furthermore the year 2014 has highest number of articles (24.19%) contributed by the authors and 2013 the lowest number of articles (17.74%) has been contributed to the Webology Journal.

| Year | Single Author | | Multi- Authors | | Total Papers | Total % age |
|---|---|---|---|---|---|---|
| | Papers | % age | Papers | % age | | |
| **2013** | 3 | 16.67 | 9 | 20.45 | 12 | 19.35 |
| **2014** | 5 | 27.78 | 10 | 22.72 | 15 | 24.19 |
| **2015** | 2 | 11.11 | 10 | 22.72 | 12 | 19.35 |
| **2016** | 5 | 27.78 | 7 | 15.91 | 12 | 19.35 |
| **2017** | 3 | 16.67 | 8 | 18.18 | 11 | 17.74 |
| **Total** | 18 | 100 | 44 | 100 | 62 | 100 |

**Table 1: Year wise Single vs Co Authorship Pattern**

4. **Analysis of Year wise distribution of Articles Published**

The table 2 explains the number of distribution of papers according to year wise. Every volume consists of 2 Issues and making 10 Issues total for the present study. It is shown that a total of 62 papers have been published during 2013-2017. In which maximum number of paper were published in the year 2014 which is accounted to. 15 (24.19%), whereas the minimum count of 11 papers were published in the year 2017 it is calculated about 17.74 percentages. There is also cumulative number of papers and cumulative percentage produced from the data obtained from the Webology Journal from the year 2013 to 2017.

| S. No | Year | Volume No. | Issues | No.of.Papers | % | Cum.No.of. Papers | Cum % |
|---|---|---|---|---|---|---|---|
| **1** | 2013 | 10 | 2 | 12 | 19.35 | 12 | 19.35 |
| **2** | 2014 | 11 | 2 | 15 | 24.19 | 27 | 43.55 |
| **3** | 2015 | 12 | 2 | 12 | 19.35 | 39 | 62.90 |
| **4** | 2016 | 13 | 2 | 12 | 19.35 | 51 | 82.26 |
| **5** | 2017 | 14 | 2 | 11 | 17.74 | 62 | 100 |
| **Total** | | | 10 | 62 | 100 | | |

**Table2: Year wise distribution of Number of Articles Published**

5. **Analysis of Author Productivity**

*Table.3 shows the data related to author's productivity during the period of study. The total number of papers* increased from 12 to 15 in the years 2013 to 2014. It gradually decreased in the next two years i.e. 2015 and 2016. Yet 2017 witnessed further decline. On the other hand the number of authors increased on par with the number of articles. Even though the number of articles calculated 31 in 2014, the average author per paper is highly shown 2.58 in this year 2015. The total average number of authors per paper observed is 2.27 and the average productivity per author calculated is 0.44. The highest number of author's productivity found in the study was 31 (0.48%) in the year 2014. The minimum number of author's productivity noted was 27 (0.41%) in the year 2017.

| Year | Total No. of Papers | Total No. of Authors | Average author/Paper | Productivity Per Author |
|------|------|------|------|------|
| 2013 | 12 | 27 | 2.25 | 0.44 |
| 2014 | 15 | 31 | 2.07 | 0.48 |
| 2015 | 12 | 31 | 2.58 | 0.39 |
| 2016 | 12 | 25 | 2.08 | 0.48 |
| 2017 | 11 | 27 | 2.45 | 0.41 |
| Total | 62 | 141 | 2.27 | 0.44 |

**Table 3: Year Wise Author Productivity**

6. **Degree of Collaboration**

Table 4 and figure 4.1 shows the degree of author collaboration analyzed in the study undertaken. To determine the extent of research productivity based on the formula given by K. Subramanyam is used. The formula is:

$$C = N_M / (N_M + N_S)$$

**C** = it represents Degree of Collaboration

$N_M$ = it represents Number of Multi – authored papers

$N_S$ = it represents Number of single authored papers

It is found that the degree of author collaboration in the Webology ranged from 0.58 to 0.83 during the period under study. Hence, C= 44 / (44 + 18) and the average value of C is = 0.71 .Therefore, as per the formula the degree of collaboration in Webology journal is 0.71.

| Year | Single Authored | Multi - Authored | DC |
|------|------|------|------|
| 2013 | 3 | 9 | 0.75 |
| 2014 | 5 | 10 | 0.67 |
| 2015 | 2 | 10 | 0.83 |
| 2016 | 5 | 7 | 0.58 |
| 2017 | 3 | 8 | 0.73 |
| **Total** | **18** | **44** | **0.71** |

**Table 4: Degree of Collaboration**

7. **Subject wise Distribution of Contributions**

The articles covered in Webology journal have been analysis on the basis of their subject coverage during the study period. The highest coverage of subject included in the journal is social networking, web 2.0 and library 2.0 and Scientometrics / Bibliometrics / Altmetrics / Webometric analysis consisting of 22.58% and 20.97% respectively. The less concentrated subject of publication brought out by this journal are about Information Seeking Behavior, ICT, Internet/Email (3.23%) which is accounted to just 2. The major count of subject of social networking, web 2.0 and library 2.0 was calculated to 14 articles followed by Scientometrics / Bibliometrics / Altmetrics / Webometric analysis consists of 13 articles. It is clear that the core concentration given by Webology journal is social networking, web 2.0 and library 2.0.

| Subjects | Numbers of Article | | | | | Total | % age |
|---|---|---|---|---|---|---|---|
| | 2013 | 2014 | 2015 | 2016 | 2017 | | |
| Social Networking / Web 2.0 / Library 2.0 | 3 | 4 | 4 | 2 | 1 | 14 | 22.58 |
| Scientometric / Bibliometric / Altmetrics /Webometric analysis | 2 | 2 | 3 | 3 | 3 | 13 | 20.97 |
| Knowledge Management | | 2 | 2 | 1 | 2 | 7 | 11.29 |
| Information Seeking Behavior | 1 | | | 1 | | 2 | 3.23 |
| E -Publishing / E - Resources | 2 | 1 | | 1 | | 4 | 6.45 |
| ICT | | 1 | 1 | | | 2 | 3.23 |
| Internet / Email | | 1 | 1 | | | 2 | 3.23 |
| Open Access | | | | 3 | 2 | 5 | 8.06 |
| Librarianship/Library Services | 1 | | | 1 | 2 | 4 | 6.45 |
| Miscellaneous | 3 | 2 | 1 | 1 | 2 | 9 | 12.9 |
| Total | 12 | 13 | 12 | 13 | 12 | 62 | 100 |

**Table 5: Subject Wise Distribution**

8. **Citation wise Distribution**

Table 6 shows that the citations were appeared at the end of contributions during 2013-2017. The highest number of contributions with citation of 11-20 were 19(31.15%) and lowest number of contributions with citation of 51-601(1.64%).

| No. of Citations | 2013 | 2014 | 2015 | 2016 | 2017 | Total | % age |
|------------------|------|------|------|------|------|-------|-------|
| 1-10 | 2 | 3 | 3 | 3 | 2 | 13 | 21.31 |
| 11-20 | 7 | 4 | 2 | 2 | 4 | 19 | 31.15 |
| 21-30 | | 3 | 2 | 4 | 1 | 10 | 16.39 |
| 31-40 | 1 | | 3 | 3 | 2 | 9 | 14.75 |
| 41-50 | 1 | 1 | 1 | | 1 | 4 | 6.56 |
| 51-60 | | 1 | | | | 1 | 1.64 |
| >61 | 1 | 2 | 1 | | 1 | 5 | 8.20 |
| Total | 12 | 14 | 12 | 12 | 11 | 61 | |

**Table 6: Citation wise Distribution**

9. **Cumulative Citation**

Table 7 represents the over view of citations published in Webology journal. Based on the analysis, the majority of 436 citations published in volume 11 in the year 2014 and occupied first place followed by 414 citations were produced by volume 14 in 2017 stood at the second place. The lowest number i.e. 253 citations with least rank volume 13.

| Year | Volume | Rank (X/Y) | Papers (Y) | Citations (X) | Average No. of Citation per Paper | Cumulative | |
|------|--------|------------|------------|---------------|-----------------------------------|-----------|---|
| | | | | | | Citations | % age |
| 2013 | 10 | 4 | 12 | 297 | 24.75 | 297 | 16.48 |
| 2014 | 11 | 3 | 15 | 436 | 29.07 | 733 | 40.68 |
| 2015 | 12 | 2 | 12 | 402 | 33.50 | 1135 | 62.99 |
| 2016 | 13 | 5 | 12 | 253 | 21.08 | 1388 | 77.03 |
| 2017 | 14 | 1 | 11 | 414 | 37.64 | 1802 | 100 |
| Total | | | 62 | 1802 | 29.06 | 1802 | 100 |

**Table 7: Cumulative Citation**

## 10. Country wise Distribution of Articles

Table 8 represents the details about country wise distribution of articles in which the maximum articles (at one rank) were contributed by authors from Iran (37.10 %), followed by USA (11.29%) followed by Russia (9.68) and then India with (8.06%) .There are three countries i.e. Estonia, Italy and Ukraine which stands at rank fifth. The Sixth Rank is for three countries i.e. Australia, London and Netherlands and six countries got number seventh rank contributing one article each.

| Rank | Country | 2013 | 2014 | 2015 | 2016 | 2017 | Total | % age of Records |
|------|---------|------|------|------|------|------|-------|------------------|
| 1 | **Iran** | 3 | 5 | 2 | 7 | 6 | 23 | 37.10 |
| 2 | **USA** | 1 | 2 | 1 | 2 | 1 | 7 | 11.29 |
| 3 | **Russia** | 1 | 2 | 2 | 1 | | 6 | 9.68 |
| 4 | **India** | 1 | | 2 | | 2 | 5 | 8.06 |
| 5 | **Estonia** | | 1 | 1 | 1 | | 3 | 4.84 |
| 5 | **Italy** | | 1 | 2 | | | 3 | 4.84 |
| 5 | **Ukraine** | | 2 | 1 | | | 3 | 4.84 |
| 6 | **Australia** | 2 | | | | | 2 | 3.23 |
| 6 | **London** | 2 | | | | | 2 | 3.23 |
| 6 | **Netherlands** | | 1 | | | 1 | 2 | 3.23 |
| 7 | **Germany** | | 1 | | | | 1 | 1.61 |
| 7 | **Macau** | | | 1 | | | 1 | 1.61 |
| 7 | **Portugal** | | | | | 1 | 1 | 1.61 |
| 7 | **Serbia** | 1 | | | | | 1 | 1.61 |
| 7 | **Spain** | | | | 1 | | 1 | 1.61 |
| 7 | **Syria** | 1 | | | | | 1 | 1.61 |
| | **Total** | 12 | 15 | 12 | 12 | 11 | 62 | |

**Table 8: Country wise Distribution of Articles**

## 11. Institution wise Contributors

Researcher has analyzed the trends of research outputs based on the institution wise contribution. Table 9 and figure 3 identifies that out of 62 papers, the highest distribution i.e. 51(82.26%) have been contributed by the academic institutions such as colleges and universities followed by 7(11.29%) and 4(6.45%) were by Special institutions & others and Research organizations respectively.

| Institution | Total | % age |
|---|---|---|
| **Academic Institutions** | 51 | 82.26 |
| **Research Institutions** | 4 | 6.45 |
| **Special Institution & Others** | 7 | 11.29 |
| **Total** | 62 | 100 |

**Table 9: Institution wise Contributors**

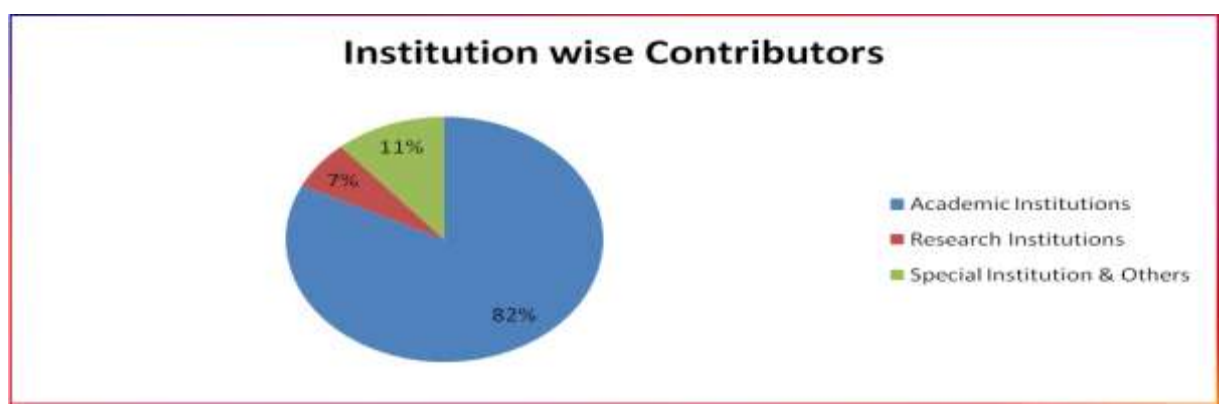

**Figure 3: Institution wise Contributors**

## V. FINDINGS

The maximum numbers of scholarly articles were published in 2014 whereas the minimum numbers of articles were in 2017. The highest numbers of papers were contributed by multi authors whereas the remaining papers were produced by single authors.  Researchers measured the distribution of degree of collaboration over the years from 2013 to 2017 as a result the degree of collaboration in Webology journal is 0.71. The majority of 436 citations published in volume no.11 in the year 2014 and occupied first place and the lowest number of citations with least rank volume 13 i.e. (253 Nos.) in the year 2016. The maximum articles were contributed by authors from Iran (37.10 %) got first place followed by USA (11.29) followed by Russia (9.68%) and India (8.06) got fourth rank respectively. The highest distributions have been contributed by academic institutions and the least number of contributions were by special institutions & others and Research organizations respectively.

## VI. CONCLUSION

Scientometrics include identifying emerging areas of scientific research, examining the development of research over time, or geographic and organizational distributions of research. Present study explored the scientometric analysis of Webology journal for the period between 2013 and 2017. The study revealed that the highest numbers of papers were contributed by multi authors whereas the remaining papers were produced by single authors. The maximum numbers of scholarly articles were published in 2014 whereas the minimum numbers of articles were in 2017.